# Is Physics Sick?

# [In Praise of Classical Physics]

# Hisham B. Ghassib

Hisham Ghassib
ghassib@psut.edu.jo
The Princess Sumaya University for Technology (PSUT),
P.O.Box 1438, Al-Jubaiha 11941, Jordan

## ABSTRACT

In this paper, it is argued that theoretical physics is more akin to an organism than to a rigid structure. It is in this sense that the epithet, "sick", applies to it. It is argued that classical physics is a model of a healthy science, and the degree of sickness of modern physics is measured accordingly. The malady is located in the relationship between mathematics and physical meaning in physical theory.

# Introduction

Increasingly, and as a concerned physicist delves deeper into modern physical theory, and exerts a genuine effort to understand what is going on in theoretical physics, he or she cannot help thinking that something is deeply amiss in modern physics; that it is afflicted with some sort of a malady or sickness that could threaten its very credibility. It is a feeling that could not be dispelled easily. One smells corruption in the air of modern theoretical physics. But, of course, this feeling implies that theoretical physics is some sort of a socio--epistemological organism. Are we justified-- logically, conceptually and historically-- in talking about physics in this strain, as though it were an organism? If so, where is the malady located? In what sense is modern physics sick? And, what is the cure?

In this paper, we argue that physical theory is indeed an organism -- a socio-epistemological organism -- that is born, and that lives, flourishes, and declines under various internal and external conditions. We, then, focus our attention on the relationship between mathematics and physical meaning in physical theory, and argue that classical physics is a paragon of health in this regard. In contrast, modern physical theory, particularly general relativity and quantum mechanics, suffers from a serious disjuncture between mathematics and physical meaning, which opens it to a barrage of absurdities and pre-scientific mystifications. This malady is not showing any sign of abating. On the contrary, it is growing at an alarming rate, which threatens the efficacy and the very existence of theoretical physics as a rational enterprise. The paper ends with a note of warning to the theoretical physics community to ponder this malady and effect fundamental changes in their methodology before the situation gets out of hand (Lindley, 1994).

## Physics as Organism

The fundamental question that pertains to the nature of physical theory is the question concerning the interconnections between physical theories in the modern era. Are physical theories a heap of independent thought systems, each with a unique structure reminiscent of a work of art? Are they free creations of the mind; leaps of creativity that emanate from a semi-rational subterranean realm of consciousness? Are they closed conceptual universes, each with a unique logic of its own?

The prevalent trend in modern physics and 20$^{th}$ century philosophy of science tends to answer the above questions in the affirmative. Physicists and philosophers of science, as varied and wide apart as the positivists (Reichenbach, 1970; Reichenbach, 1958), Karl Popper (Popper, 1986), Thomas Kuhn (Kuhn, 1970), Paul Feyerabend (Feyerabend, 1995), Gaston Bachelard (Bachelard, 1984), Werner Heisenberg (Heisenberg, 1966; Heisenberg, 1958)), and Stephen Hawking (Hawking, 1988), seem to share many views pertaining to these questions. To them, modern physical theories are not organically related to each other, but are at best contingently related. The similarities between modern physical theories are dismissed as being superficial, verbal - merely verbal – similarities. Each theory stands on its own, and comes about (I have deliberately avoided the word, "emerges") as a result of a creative irrational leap, be it individual (Popper) (Bhaskar, 1978) or social (Kuhn) (Ghassib, 1994a). It is an eternal present, unrelated either to its past or its future. The fundamental concepts that are common to the various theories, such as: mass, particle, field, Lagrangian, and Hamiltonian, are, in fact, nominally common; only the names, not the conceptual contents, are common. Thus, modern physical theories are merely a collection of essentially contingently related thought systems and ideas.

In opposition to the current wisdom regarding the nature of physical theory, we propose a different approach, which views modern theoretical physics ever since its emergence during the 17$^{th}$ century from the heart of medieval metaphysics, as an organism, as a contradictory open totality, which evolves and develops, ultimately via leaps and revolutions. According to this dialectical approach (Ghassib, 1994b), modern physical theories are dialectically related to each other, which means that they are related via necessary transformational objective relations. The In a previous publication (Ghassib, 1988), I located the fundamental objective contradiction, that has animated physics since its inception in the 17$^{th}$ century, in the contradiction between the classical concepts of particle and field. Initially, this contradiction manifested itself in embryonic form in the two self-contradictory concepts of ether and action-at-a-distance (Palter, 1960). These were embryonic substitutes for the fully developed concept of field. However, the moment the latter emerged in its fullness in the second half of the 19$^{th}$ century, it had to rip classical physics apart and expose it to a deep revolutionary crisis. This fundamental contradiction was partially resolved by special relativity (Miller, 1986; Miller, 1977), Einstein's photon theory of light (Einstein, 1905; Jammer, 1966), general relativity (D'Abro, 1950), de Broglie's theory of matter waves (Ghassib, 1983), and Schrodinger's wave mechanics (Ludwig, 1968). However, these partial resolutions have helped to reproduce this basic contradiction in various new forms, which means that it is still animating physical theory and its development. This implies that modern physical theories are not various creative, basically unrelated, conceptual systems created to cope with data and unify them. They are closer to being ways to cope with contradictions lying at the heart of prevalent theories; ways to transform existing theories into more comprehensive theories by resolving the contradictions of the former. It is in this sense that we have come to consider physical theory an organism, rather than a collection of ideas. Since physical theories emerge from each other and are transformed into each other, they cannot be severed from each other, and, thus, constitute an organic whole, a developing organism, an open totality (Ghassib, 1992).

Hitherto, two general methods have been invariably followed to resolve the basic contradictions of physical theory: (i) the reductionist method, (ii) the dialectical method. The reductionist method has invariably failed and reached an impasse, whilst the dialectical method has proven to be very fruitful, even though it has not yet totally resolved the basic contradiction. The reductionist method consists in reducing particles to fields, or vice' versa. At the turn of the $20^{th}$ century, this was followed by a number of physicists who attempted to reduce classical particles to electromagnetic fields. Amongst them were J.J. Thomson (Ghassib, 1988), Oliver Heaviside (Ghassib, 1988), Wilhelm Wien (Ghassib, 1988), Max Abraham (Ghassib, 1988), and Lorentz. Their failure led Einstein along the dialectical route on to special relativity. However, Einstein himself was later to fall into the reductionist trap in his thirty-year relentless effort to create a comprehensive unified field theory that would resolve the contradictions of general relativity (Weinberg, 1992). It was a wasted quest. Meanwhile, physical theory has evolved to forms that bear scant resemblance to Eintein's dream .

If we subscribe to this viewpoint and consider physical theory a socio - epistemological organism, we are bound to consider the question of the state of health of physical theory a legitimate question. We shall approach this question by focusing on the role of mathematics in physical theory, and, particularly, on the relationship between mathematics and physical meaning in the major physical theories of the modern era .

## Classical Physics

In classical physics, the relationship between mathematics and physical meaning is transparent and straight-forward. The essential point to notice in this regard is that classical physics uses mathematics to express physical and related philosophical concepts and meanings, which means that the physical meaning precedes the mathematical expression. The mathematics is part of the physical meaning, and not the other way round. For example, force is given a specific meaning based on a broad conception of the material world. It is viewed as a measure of material interaction between fundamental particles (atoms), and as the primary cause of change in the motion of atoms, and, thus, of all phenomena in the universe. Its properties are empirically determined, and mathematics is then used to unify and give expression to these properties (Cohen, 1983). The physical meaning is given mathematical expression because the former is fundamentally quantitative. The mathematics is not inseparable from the physical meaning; it is a necessary component of, and tool for constructing, the meaning. Newtonian mechanics is, therefore, a healthy coherent theory; a clear and distinct embodiment of a specific rationality and reason. It is based on a clear-cut family of philosophies of nature and a special broad conception, and it consists of a network of meanings and meaning systems, which underlies theoretical and empirical practices. The mathematics is part and parcel of this network and its scientific specificity.

The other disciplines of classical physics, which were modeled and based on classical mechanics, are no less clear, distinct and meaningful. The electric and magnetic fields had been discovered and physically defined before Maxwell formulated his differential equations for the electromagnetic field (Jammer1980). The latter are mathematical expressions of physical relations between physically well-defined entities. Likewise with thermodynamics and the theory of matter. In classical physics, the equations come as the climax of a process of meaning and knowledge construction, the basic tools of which are mathematics and measurement (experiment). Thus, classical equations are the highest stage of physical meaning construction. Where they are used to investigate new phenomena, they are indeed used as a tool for knowing, understanding and explaining these phenomena – i.e., for determining their essential structures, causes and mechanisms of generation. Their use is essentially a search for their concrete manifestation in the phenomena. This process deepens our knowledge of the equations and sheds more light on their efficacy, extent and significance. Thus, theory and phenomenon are dialectically related. However, there is an asymmetry in this relationship. It is the theory that is used to understand the phenomenon, and not the other way round. The theory is enriched and its meaning deepened and concretized by this process, but the process is not essential for conferring meaning on the theory, or, for gaining a basic understanding of it. That explains the perfect balance between theory and experiment in classical physics. Theory grows in parallel with experiment. The element of collaboration, coordination and cooperation between them is more pronounced than the element of antagonistic imbalance. Classical physics is indeed a sane, rationalist enterprise.

## Special and General Relativity

Modern physics (relativity and quantum mechanics) is an altogether different story. The balance, the relative harmony, hitherto encountered in classical physics, is violently severed in modern, $20^{th}$ century, physics. Now, the equation precedes the physical meaning. It is discovered prior to physical meaning, using very abstract mathematical arguments and relations. The search for meaning starts after the discovery of the equation and continues seemingly indefinitely. Experiment, philosophy and hypothesis are all utilized to discover the missing physical meaning. The only objective reality practically recognized is the uninterpreted mathematical equation; a clear return to Pythagorean-Platonic thinking (Heisenberg, 1971). The equation is the only objective fact. All else is subjectivity.

Needless to say, it all starts with the maestro of modern physics, Albert Einstein. This Pythagorean trend is evident right from the start, even in special relativity, that indispensable tool of modern investigation. The special retativistic transformation (Lorentz) equations are first derived, and then given a number of physical interpretations. The controversy, of course, is not yet over. The energy-mass equation is first mathematically derived, using the solid methods of classical mechanics, and later on interpreted physically. A mathematical, four-dimensional, space is constructed (Minkowski's spacetime), and its mathematical properties and symmetries are used to reconstruct classical mechanics special relativistically. The physical interpretation comes later, of course not without the help of the "discarded'' classical mechanics. Thus, modern physics is mathematics in search of physical meaning. It has indeed been reduced to a branch of mathematics.

What is a barely noticeable trend in special relativity assumes enormous proportions and the status of a fundamental principle in Einstein's general relativity -- namely, the principle of covariance. The coordinates lose all specific physical meaning, and the laws of physics are constructed out of tensors on a four-dimensional manifold in which real space and real time are dissolved beyond recognition. The abstract engulfs the concrete and consumes it beyond retrieval. Being covariant, the laws of physics are constructed in such a way that they are impervious and indifferent to the choice of coordinate system. In constructing his field equations, Einstein was guided as much by mathematical considerations as by physical grounds. His search focused on finding a tensor which satisfied certain mathematical (geometric) criteria, which he then equated with an ambiguous generalization of classical mass and energy (energy - momentum tensor)(Schilpp, 1951). Admittedly, the Einstein field equations did (and do) have a broad geometrico-physical meaning embedded in them-- namely, spacetime curvature is related to matter and motion. However, this apriori assignation of meaning is rather deceptive. When it comes to solving the Einstein field equations, one is immediately confronted with the choice of a suitable coordinate system and the physical interpretation of the system chosen (Fok). Needless to say, recourse in invariably made to classical mechanics to interpret certain fundamental constants in the theory (e.g., Newton's universal gravitational constant, G, and the mass density) (Eddington, 1965). Also, finding a suitable coordinate system and interpreting it reasonably are no straight-forward matter. It is a very laborious process which relies heavily on the clear and distinct ideas of classical physics. This is most clearly evident in the history of the Schwarzschild solution to Einstein's field equations (Israel, 1987). It took the best minds of the scientific community more than thirty years after Schwarzschild's discovery to arrive at a "natural" coordinate system (the Kruskal-Szekeres system)(Misner, 1971), and a tolerably reasonable interpretation of it. Meanwhile, the whole matter led to such real and apparent absurdities that no one working in the field at the time, least of all Einstein and Eddington, made heads or tails (apart from the mathematics, of course) of what one was actually doing. However, even after the major discoveries of Kruskal, Szekeres, Kerr, Wheeler, Penrose, Hawking, Zeldovich, Novikov, Ellis, Israel and others (Misner, 1971), the majority of theoretical physicists still feel rather lost when it comes to physically understand black hole and wormhole physics. The mathematics is pretty clear, though heavy, but the physics is lost in the mathematical entanglement. In spite of the apparent experimental successes of general relativity, one is left with an uneasy feeling of intangibility and contrivance (Dunbar, 1996).

Contrary to the myth propagated by a number of 20th century philosophers of science, general relativity is not a "beautiful" self-contained and self-sufficient theory -- a finished work of art (Bachelard, 1984). Rather, and like all other physical theories, it is a contradictory open totality, which is heavily dependent on classical theory in both form and content. In a sense, it is a field (as opposed to particle) generalization of classical theory (Einstein, 1961; Einstein, 1950). Being afflicted with the malady of building its mathematical equations before establishing physical meanings, it is bound to be heavily dependent on classical physics in its search for physical meaning.

## Quantum Mechanics

This malady reaches absurd proportions in quantum mechanics. Once again, a myth has been propagated, by both theoretical physicists and philosophers of science, to the effect that quantum mechanics is a complete theory, which has solved the contradictions, and overcome the inadequacies, of the old quantum theory. But, has it really done that? Is it less inadequate and contradictory than the old quantum theory? Is it not more prudent to recognize its incompleteness and inadequacies, and look for ways to "develop" it back towards, rather than away from, the great insights of classical physics?

Instead, we have seen leading theoretical physicists trying to enthrone quantum mechanics and elevate it to the status of "the" complete theory of physics (Cushing, 1994). Of course, to conceal its glaring inadequacies, they have had to invite all sorts of absurd ideas and interpretations. In the process, they have opened large holes in the edifice of physics for all sorts of pre-scientific and irrationalist, almost solipsistic, nonsense. How else would a rationalist describe the Copenhagen assertion that the conscious observer (consciousness) creates external material reality (Mermin, 1985; Popper, 1982), that the humanly willful and purposeful act of measurement creates microscopic reality (Landau, 1965), that each event splits the universe into many universes (Everett III, 1957), or, that the present fixes the past (Wheeler, 1957). These absurdities satirized by Schrodinger with his famous cat (Schrodinger, 1935; Beller, 1977), are, in fact, rooted in the aforementioned malady of modern physics -- namely, that the mathematics precedes the physical meaning and conditions it; that the latter is a mere aftereffect or epiphenomenon.

When Schrodinger discovered his equation, he thought that he was discovering a new type of classical field, and that its nature would be specified later in a manner similar to the way the optical wave–field, discovered by Huygens, Young and Fresnel, was later shown to be Maxwell's electromagnetic field (Harman, 1995). Of course, things have not developed in this way, and, already, almost ninety years have passed since the construction of quantum mechanics, and no convincing interpretation seems to be in sight (Squires, 1996).

This state of affairs applies equally to the other formulations of quantum mechanics. Heisenberg started with an insistence on observables (D'Abro, 1952), and ended up endorsing a form of solipsism, whereby the laws of quantum physics do not reflect physical reality, but a mental state (our possible knowledge) (Heisenberg, 1960; Heisenberg, 1961). The same dilemma recurs in Wigner's formulation of quantum mechanics (Wigner, 1932). In classical transport theory, the distribution function is first defined and assigned a specific meaning, and the Liouville equation is then derived by translating the physical meaning into mathematical operations (Jancel, 1963). Thus, the Liouville equation is a mathematical embodiment, or, rather, a development, of the physical meaning. In quantum transport theory, the opposite occurs. The Wigner distribution function is defined only mathematically, in terms of the ill-defined Schrodinger wave function, and the Schrodinger equation is used to derive the quantum Wigner equation (Ghassib, 1996). The never-ending search for physical meaning starts after the equation is specified. Needless to say, this search is principally guided by the "discarded" classical Liouville equation. Once again, we have an equation in search of physical meaning (Planck, 1933; Lewontin, 1993). We have a fictional phase space unrelated to real events, and we have a physically indefinite distribution function mathematically defined on it. The latter seems to be a mere calculational device devoid of physical meaning.

**Conclusion**

The foregoing analysis raises a number of urgent questions which should be confronted by modern theoretical physics. Can it be that the very methods of modern theoretical physics, as opposed to classical theoretical physics, are at fault? Is it rationally permissible to construct equations for physically undefined, ill-defined or incomprehensible mathematical constructs? Was the path connecting classical physics to modern physics inevitable -- i.e., written in the heart of the theoretical physics organism? Or, could this sickness have been avoided? Was it the result of an ideological intervention that managed to deflect physics from its true, healthy path? Should modern theoretical physicists not slow down their feverish, frenzied competition to ponder, and rethink, the methods they have been employing for the last ninety years? Should they not revise these methods and their results in relationship to classical theoretical physics? Should they not carry out a rigorous ideological analysis of their enterprise, and learn the critical methods of the social sciences? Should they not relearn how to think?[53]

We tend to answer these questions in the affirmative, and we leave them standing as a warning sign to the theoretical physics community.

In future publications, we intend to develop the principal ideas propounded in this paper in relation to recent developments in quantum gravity, superstring theory and quantum cosmology.